\documentclass[aps,showpacs,preprintnumbers,amssymb,superscriptaddress,twocolumn]{revtex4}

\usepackage{graphicx,color}
\usepackage{dcolumn}
\usepackage{bm}
\usepackage[active]{srcltx}

\begin{document}
\title{Hidden quantum phase transition in Mn$_{1-x}$Fe$_{x}$Ge: evidence brought by small-angle neutron scattering.}

\author{E. Altynbaev}
\affiliation{Petersburg Nuclear Physics Institute, Gatchina, 188300 St-Petersburg, Russia}
\affiliation{Faculty of Physics, Saint-Petersburg State University, 198504 Saint Petersburg, Russia}
\author{S.-A. Siegfried}
\affiliation{German Engineering Materials Science Centre (GEMS) at Heinz Maier-Leibnitz Zentrum (MLZ), Helmholtz-Zentrum Geesthacht GmbH, Lichtenbergstr. 1, 85747 Garching bei München, Germany}
\author{E. Moskvin}
\affiliation{Petersburg Nuclear Physics Institute, Gatchina, 188300 St-Petersburg, Russia}
\affiliation{Faculty of Physics, Saint-Petersburg State University, 198504 Saint Petersburg, Russia}
\author{D. Menzel}
\affiliation{Technische Universit\"at Braunschweig, 38106 Braunschweig, Germany}
\author{C. Dewhurst}
\affiliation{Institut Laue-Langevin, 6 Rue Jules Horowitz, F-38042 Grenoble, France}
\author{A. Heinemann}
\affiliation{Helmholtz Zentrum Geesthacht, 21502 Geesthacht, Germany}
\author{A. Feoktystov}
\affiliation{Jülich Centre for Neutron Science (JCNS) at Heinz Maier-Leibnitz Zentrum (MLZ), Forschungszentrum Jülich GmbH, Lichtenbergstr. 1, D-85748 Garching, Germany}
\author{L.Fomicheva}
\affiliation{Institute for High Pressure Physics,  142190, Troitsk, Moscow Region, Russia}
\author{A. Tsvyashchenko}
\affiliation{Institute for High Pressure Physics,  142190, Troitsk, Moscow Region, Russia}
\author{S. Grigoriev}
\affiliation{Petersburg Nuclear Physics Institute, Gatchina, 188300 St-Petersburg, Russia}
\affiliation{Faculty of Physics, Saint-Petersburg State University, 198504 Saint Petersburg, Russia}

\begin{abstract}

The magnetic system of the Mn$_{1-x}$Fe$_{x}$Ge solid solution is ordered in a spiral spin structure in the whole concentration range of $x \in [0 \div 1]$. The close inspection of the small-angle neutron scattering data reveals the quantum phase transition from the long-range ordered (LRO) to short range ordered (SRO) helical structure upon increase of Fe-concentration at $x \in [0.25 \div 0.4]$. The SRO of the helical structure is identified as a Lorentzian contribution, while LRO is associated with the Gaussian contribution into the scattering profile function. The scenario of the quantum phase transition with $x$ as a driving parameter is similar to the thermal phase transition in pure MnGe. The quantum nature of the SRO is proved by the temperature independent correlation length of the helical structure at low and intermediate temperature ranges with remarkable decrease above certain temperature $T_Q$. We suggest the $x$-dependent modification of the effective Ruderman-Kittel-Kasuya-Yosida exchange interaction within the Heisenberg model of magnetism to explain the quantum critical regime in Mn$_{1-x}$Fe$_{x}$Ge.

\end{abstract}


\maketitle

The cubic B20-type compounds (MnSi, etc) are well known for the incommensurate magnetic structures with a very long period appeared due to noncentrosymmetric arrangement of magnetic atoms. It is widely recognized that the helix spin structure is built on the hierarchy of interactions: ferromagnetic exchange interaction, antisymmetric Dzyaloshinskii-Moryia interaction (DMI), and the anisotropic exchange interaction \cite{Kataoka, Bak}. It is also known that the substitution of manganese by iron in the isostructural solid solutions Mn$_{1-x}$Fe$_x$Si suppresses the helical spin state \cite{Nishihara}. The neutron scattering studies \cite{Grigoriev_09_PRB, Grigoriev_11_PRB1} together with magnetic data and specific heat measurements \cite{Nishihara, Bauer_10_PRB, Demishev_JETP_2014} discovered a quantum critical point (QCP) corresponding to the suppression of the spin spiral phase with long-range order (LRO) in Mn$_{1-x}$Fe$_x$Si. This QCP located at $x_{c1} \approx 0.11 - 0.12$ is, however, hidden by a short-range order of the spin helix (SRO) \cite{Grigoriev_11_PRB1, Bauer_10_PRB, Demishev_JETP_2014} that agrees well with the theoretical models \cite{Tewari_PRL_2006, Kruger_PRL_2012}. This SRO phase, sometimes referred as chiral spin liquid \cite{Tewari_PRL_2006}, which is destroyed at the second QCP $x_{c2} \approx 0.24$. Thus it has been shown that Mn$_{1-x}$Fe$_x$Si undergoes a sequence of the two quantum phase transitions \cite{Demishev_JETP_2014}.

The real breakthrough in understanding of the experimental facts mentioned above has been done via scrutinizing the Hall effect in Mn$_{1-x}$Fe$_x$Si \cite{Demishev_PRL_2015}. It was found that the substitution of Mn with Fe results rather in the hole doping opposite to the natural expectations on the electron doping. The two groups of the charge carriers contribute to the Hall effect and the ratio between them changes the sign of the Hall effect constants at $x_{c1} \approx 0.11$, what is definitely associated with the QCP in these compounds. Despite the fact that the solid solutions of Mn$_{1-x}$Fe$_x$Si are often considered as itinerant magnets \cite{Tewari_PRL_2006, Kruger_PRL_2012}, recent magnetic resonance and magnetoresistance studies \cite{Demishev_PRB_2012, Demishev_JETP_2014_2} favor the alternative explanation based on the Heisenberg localized magnetic moments (LMM) model of Mn ions. Furthermore the discovered inversion of the Hall constants should results in the modulation of the effective Ruderman-Kittel-Kasuya-Yosida (RKKY) exchange interaction within the Heisenberg model of magnetism. Considering the MnSi as a DMI-based helimagnet, the role of RKKY interaction is to compete with the DMI and serve as a tool for destabilization of the helical structure at $x_{c1}$.

In this Letter we focus on the similarly hidden quantum phase transition in Mn$_{1-x}$Fe$_{x}$Ge compounds. Since the magnetic system is ordered in a spiral spin structure in the whole concentration range of $x \in [0 \div 1]$ \cite{Grigoriev13PRL}, we use the small angle neutron scattering (SANS) technique to show that the LRO is transformed into the SRO upon Mn replacement with Fe at $x \in [0.25 \div 0.4]$. The helix instability of the quantum nature dominates over the thermal spin helical fluctuations up to $T_{QF} \sim 60-90$ K in the same concentration range. The same mechanism as in \cite{Demishev_PRL_2015} is applied to explain the hidden QPT in Mn$_{1-x}$Fe$_x$Ge.

 \begin{figure}
 \includegraphics[width=0.45\textwidth]{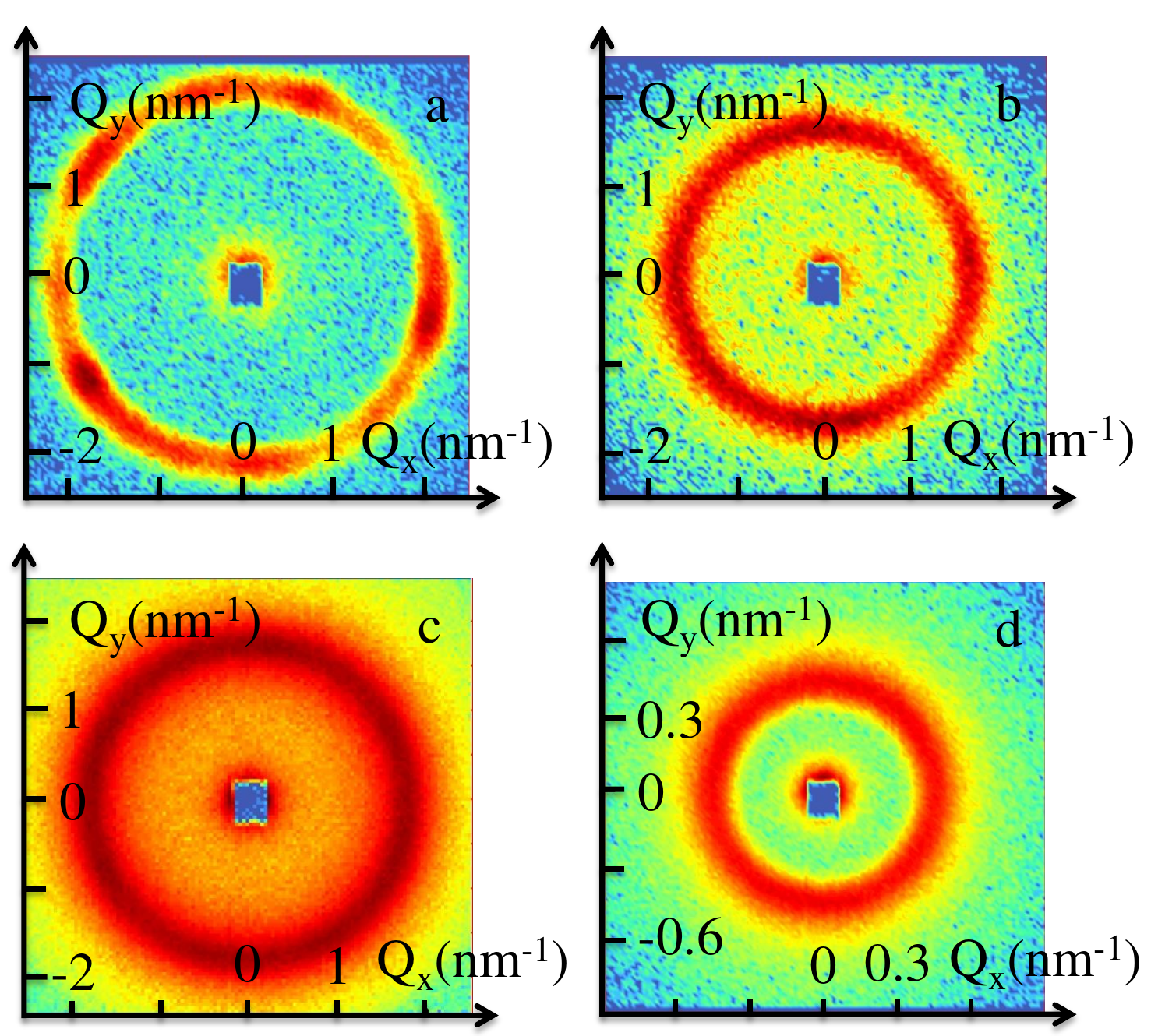}
 \caption{Examples of the neutron scattering maps for Mn$_{1-x}$Fe$_x$Ge compounds with $x = 0.0$ (a), $0.25$ (b), 0.4 (c) and 0.5 (d) at $T = 5$ K taken at zero field.}
 \label{fig:mapsx}
 \end{figure}

The Mn$_{1-x}$Fe$_{x}$Ge solid solution demonstrates intriguing magnetic properties \cite{Grigoriev13PRL, Tokura_nano_2, Lebech89, Kanazawa11, Kanazawa12, Makarova12, Alt, Mirebeau,Moskvin_PRL_2013}. It was recently shown that the helix chirality is altered by mixing the two types of magnetic atoms (Fe and Mn) on the Fe-rich side of the phase diagram \cite{Grigoriev13PRL, Tokura_nano_2}. The compounds with $x \geq 0.5$, are charaterized by the long period of the helix structure, which becomes infinite at $x_c = 0.75$, {\it i.e.} the compound transforms to ferromagnet. The change of the helix chirality at $x_c$ was also experimentally observed via the change of the sign of the DM interaction. The DM interaction is positive for compounds with $x < x_c$ and negative for compounds with $x > x_c$ \cite{Grigoriev13PRL,Tokura_nano_2}. The {\it ab-initio} calculations can reasonably reproduce the experimentally observed inversion of $D$ in the Mn$_{1-x}$Fe$_x$Ge close to the $x_c$ \cite{Koretsune_arxiv_2015,Gayles_arxiv_2015,Kikuchi_arxiv_2016}.

On contrary, the compounds of the Mn-rich side of the phase diagram possess a short period spin helix. The small angle neutron scattering \cite{Alt} and M\"ossbauer spectroscopy \cite{Mirebeau} show that the stable helical structure at $T = 0$ becomes intrinsically unstable upon temperature increase. The temperature activates both unusual spin excitations and helical spin fluctuations, which result in the phase transition to fluctuating helical state at $T_N = 130 \pm 2$ K. The heli- to paramagnetic phase transition of the pure MnGe is spread over 100 K above the critical temperature $T_N$ \cite{Alt}, which differs strongly from the scenario of the phase transition of any B20 compounds \cite{Lebech89,Moskvin_PRL_2013,Grigoriev05PRB}.

At present there is a general belief that the ratio between the ferromagnetic exchange interaction and Dzyaloshinskii-Moriya interaction should determine the value of the helix wave vector in MnGe as well as in other B20 compounds. This belief is based on the well established description (Bak-Jensen model) of the MnSi and FeGe \cite{Kataoka, Bak}, where $k_s$ is rather small and equal to 0.35 nm$^{-1}$ for MnSi and 0.09 nm$^{-1}$ for FeGe. The value of the helix wave vector for MnGe is however much larger and is equal to $k_s = 2.2$ nm$^{-1}$ while the value of the critical field, needed to transform the helimagnet to the ferromagnetic state, is equal to $H_{c2} = 15$ T. This fact could be hardly explained within the conventional Bak-Jensen model for B20 helimagnets \cite{Kataoka, Bak}. The experimental facts shown below together with relatively small predicted value of DM interaction in MnGe (at $x = 0$) \cite{Koretsune_arxiv_2015, Gayles_arxiv_2015,Kikuchi_arxiv_2016} leads to the conclusion that the spin helix in MnGe is based on the effective RKKY interaction. Due to the fact that the magnetic structure of FeGe is based on DMI, one would expect the transition from RKKY spin helix to DMI spin helix with $x$. The $x$-dependence of the helix wave vector $k$ shows that the helical structure with $k \sim 2$ nm$^{-1}$ in the range $x \leq 0.4$ is replaced by the structure with small wave vector of the helix in the range $x \geq 0.5$ meaning the value $x_{c2} \approx 0.45$ as the critical concentration for the quantum phase transition \cite{Grigoriev13PRL}.

The polycrystalline samples of Mn$_{1-x}$Fe$_{x}$Ge compounds have been synthesized by high pressure method at the Institute for High Pressure Physics, Troitsk, Moscow, Russia. As it can be only synthesized under high pressure, the sample have a polycrystalline form with a crystallite size not less than 10 microns (see \cite{Tsvyashchenko84} for details). The X-ray powder diffraction confirmed the B20 structure of the samples used in the experiments \cite{Chernyshov}. This study has not revealed a dispersion of the concentration $x$ larger than 1-2 \%. The small angle neutron scattering have shown that the spinodal decomposition with the large distribution of $x$ of order of 5-10 \% occurs only within the small fraction of the sample (similar to those studied in \cite{Tokura_nano_2}), while the most of the sample shows distribution of the $x$ not larger that 2 \%. We ascribe the fraction of the samples with relatively large $x$ distribution to the surface of the grains in the polycrystalline material. Taking into account that the diffraction technique averages over the full volume of the sample, the imperfectness of the samples can not affect the intensity profile and prevent one from the evaluation of the correlation functions.

The SANS measurements were carried out at instruments D11 (ILL, Grenoble, France), SANS-1 \cite{SANS} and KWS-1 \cite{KWS} (FRM-II reactor, Garching, Germany). Neutrons with a mean wavelength of $\lambda = 0.6$ nm were used. The sample-detector distance of $2$ m was set to cover the scattering vector range $Q$ from $0.7$ nm$^{-1}$ to $2.7$ nm$^{-1}$ with the resolution equal to $0.1$ nm$^{-1}$. The scattering intensity is measured upon zero field cooling from the paramagnetic phase at $T = 300$ K to the ordered phase at $T = 5$ K.

Figure \ref{fig:mapsx}(a-d) shows examples of the small angle neutron scattering maps for Mn$_{1-x}$Fe$_x$Ge compounds with $x$ from 0.0 to 0.5 at $T = 5$ K. The typical powder-like images were detected with anisotropic rings of intensity for samples with $x = 0.0$ and 0.2. The observed spots are referred to the scattering from the relatively large magnetic domains of the helical spin structure limited by the crystal grains sizes. The intensity distribution within the ring becomes isotropic with increase of the iron concentration meaning that the helical domains breaks into smaller pieces within the grain.

\begin{figure}
 \includegraphics[width=0.45\textwidth]{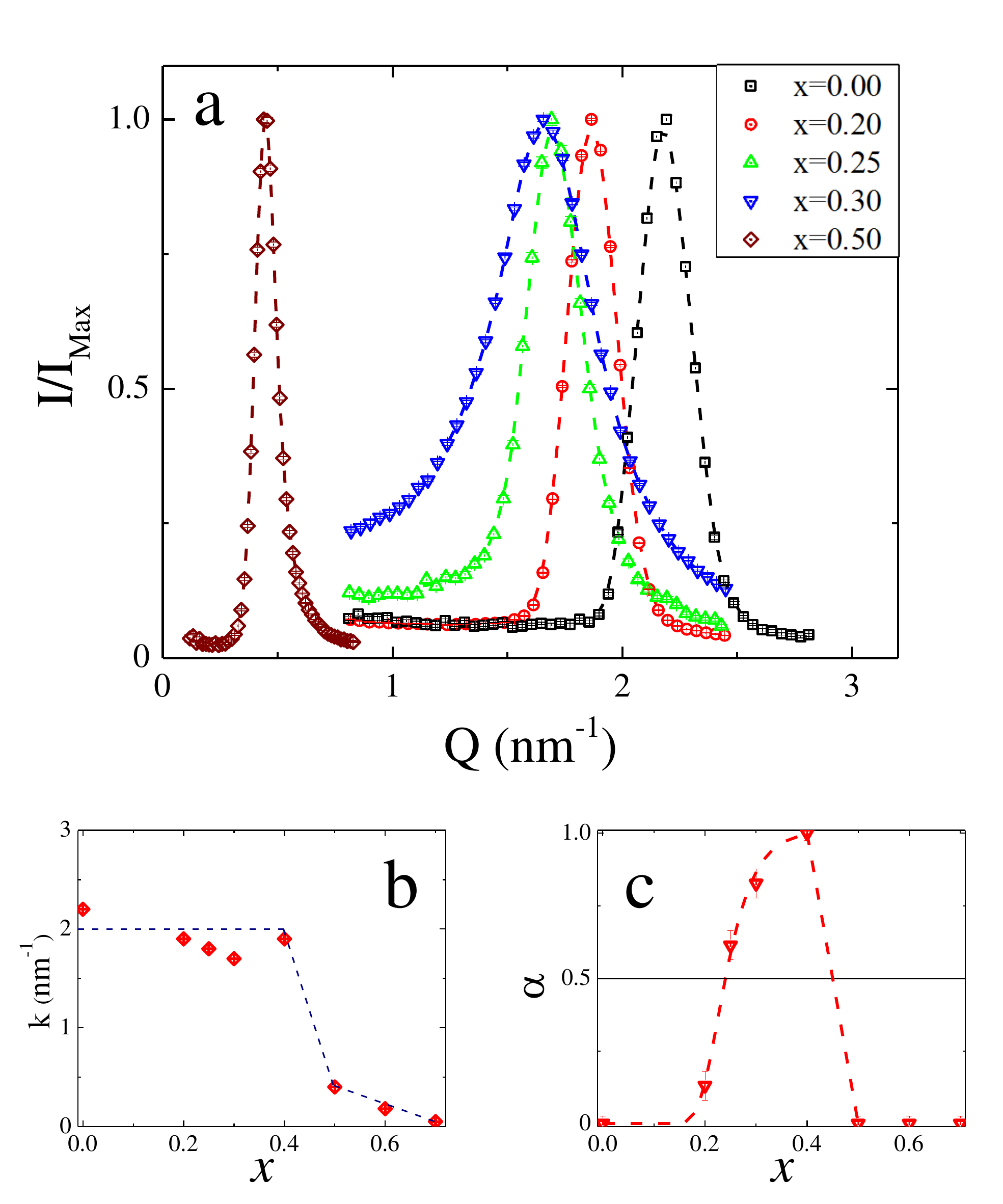}
 \caption{(a) Momentum transfer dependence of the scattering intensity at $T = 5$ K for different Mn$_{1-x}$Fe$_x$Ge compounds. (b) $x$-dependence of the helix wave vector value $k$ \cite{Grigoriev13PRL}. (c) $x$-dependence of the Lorentzian fraction $\alpha$ in the peak at $T=5$ K, which is associated to the fluctuating helical phase. Lines are the guide for the eyes.}
 \label{fig:avx}
 \end{figure}

The scattering intensity $I({\bf Q})$ measured at $T = 5$ K was azimuthally averaged and plotted in Fig.\ref{fig:avx}a. For better comparison the intensity was normalized to its maximum $I/I_{Max}$. The $x$-dependence of $k$ is presented in Fig.\ref{fig:avx}b. The scattering function (Bragg peak) of the pure MnGe can be well approximated by Gaussian (Fig.\ref{fig:avx}a). The shape of the scattering function changes upon Mn replacement with Fe ($x \in [0.2 \div 0.4]$) and can only be described by the pseudo-Voigt function with four different parameters: the scaling factor $I_{Max}$, the Lorentz fraction $\alpha$, the peak position $k$ and the width of both, Gaussian and Lorentzian functions $\kappa$. The intensity profile can be described again by the pure Gaussian for compounds with $x \geq 0.5$. The asymmetry of the peak for $x=0.5$ is referred to the spinodal decomposition of the compounds as long as the small shift of the $x$ parameter results in significant change of the $k$ value (Fig.\ref{fig:avx}b). 

\begin{figure}
\includegraphics[width=0.45\textwidth]{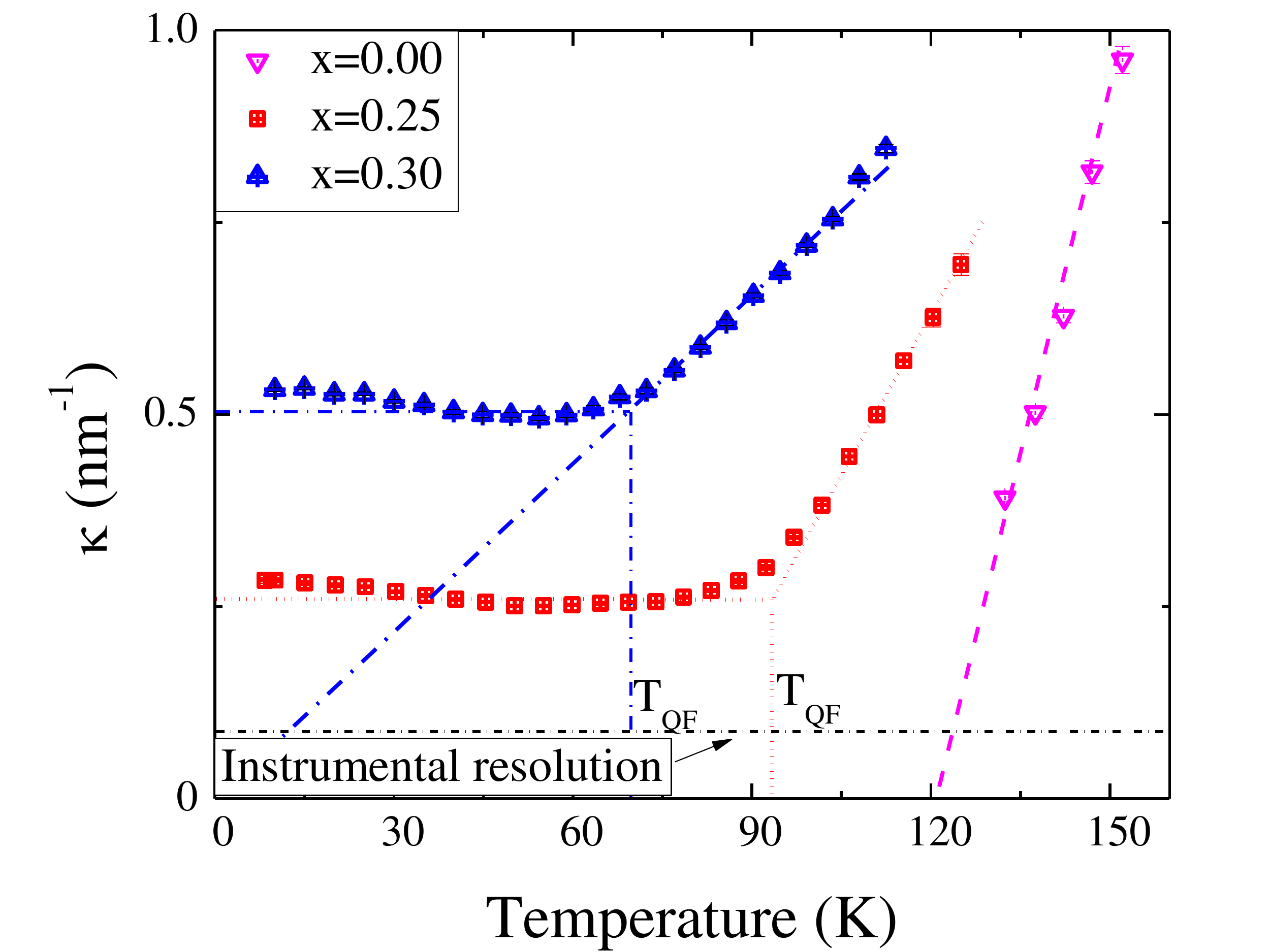}
\caption{(color online). Temperature dependence of the inverse correlation length of helical fluctuations, $\kappa =1/\xi$, for Mn$_{1-x}$Fe$_x$Ge with $x = 0.0$, 0.25 and 0.3. Lines are the guide for the eyes.}
\label{fig:kappaxt}
\end{figure}

The analysis of the temperature evolution of the scattering curves for all samples was performed in the same manner as for pure MnGe. The shape of the Bragg reflection is well described by the single Gaussian function at low temperatures and $x = 0.0$ ($\alpha = 0$). With increase of the temperature or $x$ the profile is, firstly, transformed into the pseudo-Voight function ($0 < \alpha < 1$) and, secondly, is contaminated by additional (abnormal) scattering at $Q < k$ accosiated to the inelastic scattering. Within the high temperature range or at $0.3 < x < 0.4$ the intensity profile represents the sum of the Lorentzian and abnormal scattering ($\alpha = 1$). The characteristic temperatures related to different regimes decreases smoothly with increase of the Fe concentration. 

The Lorentzian contribution into the scattering corresponds to the scattering from SRO of the helix structure, while the Gaussian contribution comes from the LRO \cite{Pat_Pok}. As the Bragg reflection is well described by the sum of Lorentzian and Gaussian functions with the same width and peak position, one can separate the fractions of the helical fluctuations and the stable helices in the compound. The helical fluctuations have to have the finite correlation length $\kappa$ and lifetime $\tau$ which are much smaller than the characteristic parameters for LRO of helical structure \cite{Grigoriev10PRB}. The $x$-dependence of the Lorentz fraction $\alpha$, which can be counted as the fraction of helical fluctuations, is presented in Fig.\ref{fig:avx}c.

The fraction of SRO dominates over the fraction of the LRO at $x > 0.25$ (Fig.\ref{fig:avx}c) showing that the LRO of the stable helix disappears and is gradually replaced by the SRO at low temperatures. Even if the SRO is ascribed to the helical fluctuations it could not be considered as the paramagnetic state of the structure. More accurately this process should be described similarly to the one observed in pure MnGe as a function of the temperature \cite{Alt}, where the LRO is gradually replaced by the SRO with the temperature in the range from 80 K to $T_N = 130$ K, while the helical fluctuations are clearly observed up to $T_h = 150$ K. 

As long as the nature of the disorder is clearly provided by the Fe replacement of Mn atoms, the origin of SRO at low temperatures can be explained by the model similar to the Mn$_{1-x}$Fe$_x$Si. In case of Si-based compounds, the Fe-doping results in increase of the hole concentration instead of electron concentration which are considered as the driving force for tuning the quantum critical regime via modifying the effective Ruderman-Kittel-Kasuya-Yosida exchange interaction within the Heisenberg model of magnetism \cite{Demishev_PRL_2015}. This model should be inverted for Mn$_{1-x}$Fe$_x$Ge compounds. The LRO of the helix structure is build on the main effective RKKY interaction and small DMI constant for the pure MnGe. The RKKY interaction decreases and DMI increases with $x$ that leads to the quantum phase transition through SRO of the helix fluctuations at $x > 0.25$. This model does not contradict to the experimental data obtained either in this work or in any others provided till present time.

Another evidence of the competition between different interactions that built helical order is the evolution of the correlation length of the structure. The estimation of the correlation length and the size of the incommensurate magnetic helix is always limited by the resolution of the SANS instrument $\xi_{max}$. For systems with the LRO ($\xi > \xi_{max}$) the width of the peak is always equal to the the instrumental resolution $\kappa_{res} = 2\pi/\xi_{max}$. For systems with the SRO, the correlation length is smaller than the instrument resolution, $\xi < \xi_{max}$ and fits ideally within the scope of the SANS instrument. The width of the peak $\kappa$ is considered as an inverse correlation length of the magnetic structure $\xi = 2\pi/\kappa$. The temperature dependence of the inverse correlation length $\kappa$ is presented in Fig.\ref{fig:kappaxt} for Mn$_{1-x}$Fe$_x$Ge with $x = 0.0$, 0.25 and 0.3.

The inverse correlation length $\kappa$ of the helical fluctuations is expected to increase with temperature close to the order-disorder magnetic phase transition meaning the decrease of the correlation length of the fluctuations. Such behavior is well seen in Fig.\ref{fig:kappaxt}. Nevertheless, the helical fluctuations are also observed at low temperatures for Mn$_{1-x}$Fe$_x$Ge with $x > 0.2$ (Fig.\ref{fig:avx}c). The correlation length $\xi = 2\pi/\kappa$ of the helical fluctuations remains constant but still smaller than the highest reachable value for the SANS instrument ($\xi < \xi_{max}$) in a wide temperature range (Fig.\ref{fig:kappaxt}). The existence of two different temperature regimes implies the different states of the magnetic system: the thermal spin helix fluctuations, which evolve with temperature and the $T$-independent type of the SRO at low temperatures. We define the crossover temperature of these two regimes as $T_{QF}$. The examples of such determination of the crossover temperatures $T_{QF}$ are presented in Fig.\ref{fig:kappaxt}. Due to the fact that the correlation length of the helical fluctuations are temperature independent at $T < T_{QF}$ we consider the fluctuations to be of the quantum nature, since there should be another reason but the temperature that decreases the correlation length of the fluctuation. If one suggests only temperature as the energy regulating the correlation length of the fluctuation, then the size of the helical fluctuation should increase infinitely with $T\rightarrow T_c$ while $T_c \rightarrow 0$ with $x$. For example, such tendency can be seen for compounds with $x = 0.3$ (Fig.\ref{fig:kappaxt}). Nevertheless, the increase of the correlation length of the helical fluctuation with decrease of the temperature appears to be limited at $T = T_{QF}$ by a certain reason of non-thermal origin. One can estimate that it limits the correlation length by the value of $l_c = 2\pi/\kappa \approx 10 $ nm, which is approximately 3 times larger than the period of the magnetic helix.

\begin{figure}
\includegraphics[width=0.5\textwidth]{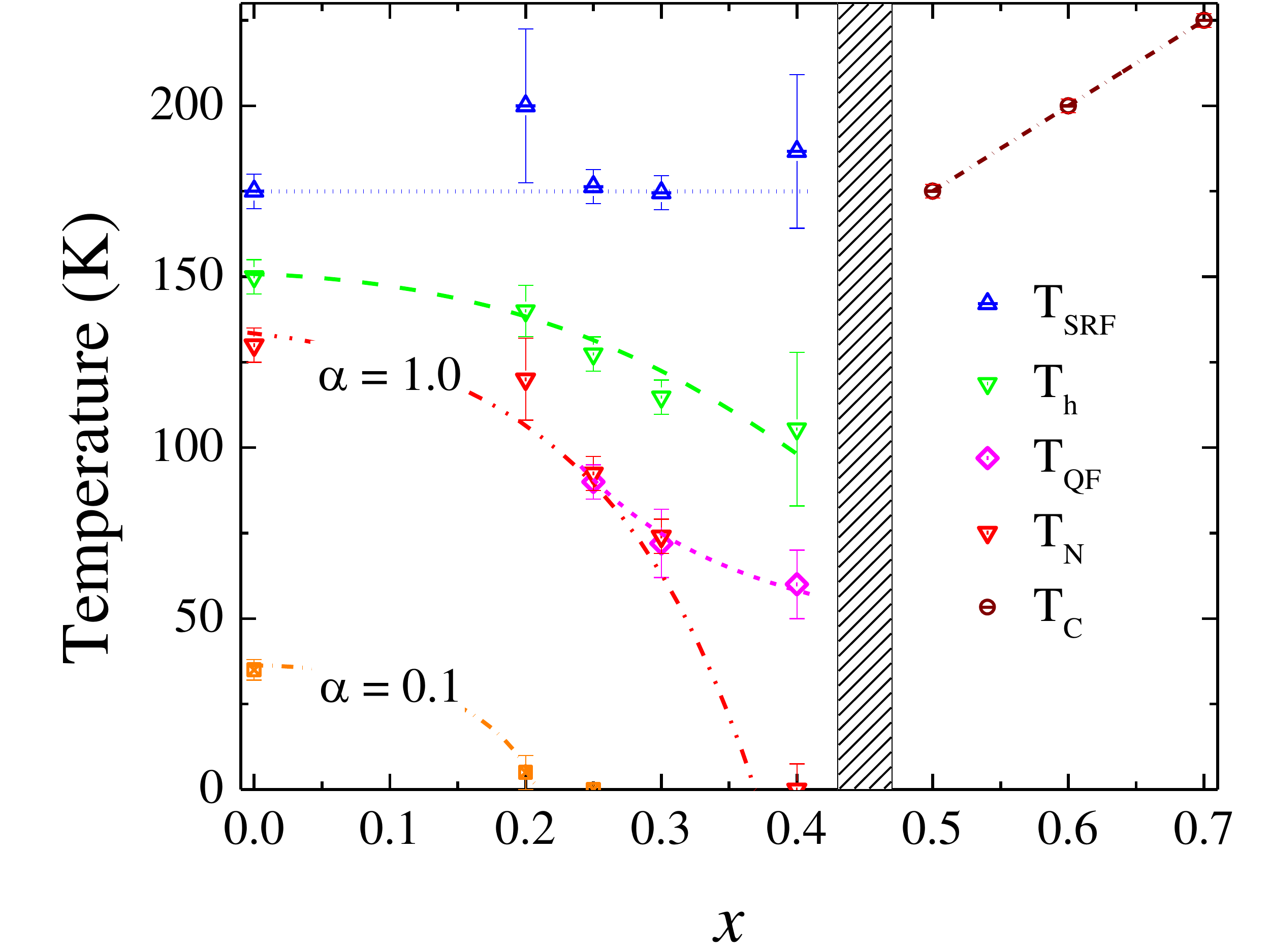}
\caption{(color online). $T-x$ phase diagram of the magnetic structure of Mn$_{1-x}$Fe$_x$Ge compounds. $\alpha$ represents the fraction of the fluctuating spiral phase. The stable spiral phase (LRO) with $\alpha <0.1$ is limited with the corresponding line in the left-down corner of the plot. The line $\alpha = 1.0$ defines the ($x-T$) value of the transition to the 100 \% fluctuating spiral state. The corresponding temperature is defined as $T_N$. The temperature $T_h$ determines the upper border of the fluctuating spiral phase. The temperature $T_{SRF}$ defines the lower border of the short-range ferromagnetic fluctuations. The temperature $T_{QF}$ defines the upper border of the quantum fluctuating state. The temperature $T_c$ is the only critical point found for the compounds with $x \geq 0.5$. The vertical band at $x \approx 0.45$ represents the transition from the RKKY-type to the DMI-type of spirals. Lines are the guide for the eyes.}
\label{fig:Temps}
\end{figure}

The ($T-x$) phase diagram of the magnetic structure of Mn$_{1-x}$Fe$_x$Ge compounds is plotted in Figure \ref{fig:Temps}. As was found in \cite{Alt}, the temperature evolution of magnetic system of the pure MnGe compound undergoes a series of crossovers from one state to another. From the analysis of the scattering function we determined three different temperatures for MnGe. The helical peak can be distinguished below $T_{h} = 150$ K. The complex mixture of the fluctuating spins, which could not be identified as a certain type of structure, was observed at temperatures $T_{h} < T < T_{SRF} = 180$ K. This mixture, nevertheless, is transformed into the ferromagnetic fluctuations (defined as short-range ferromagnetic (SRF) state) at $T_{SRF}$ with characteristic size less than 2 nm. It should be noted that the Fe-replacement in Mn$_{1-x}$Fe$_x$Ge does not affect this SRF state at the high-temperature region. The spiral state below $T_h$ consists of the fraction of the fluctuating spiral $\alpha$ and the fraction of the stable spiral $(1-\alpha)$. Nevertheless, the 100\% fluctuating spiral state occurs in the large area of the $(T-x)$ phase diagram starting with the lower border marked as a line with $\alpha = 1.0$ up to the $T_h$. The temperature corresponding to the line $\alpha = 1.0$ is defined as $T_N$. The temperature $T_h$ decreases smoothly with $x$, while the temperature $T_N$ tends to zero.

The helical fluctuations at temperatures far below $T_N$ were observed even for the pure MnGe compound \cite{Alt}. The coexistence of the LRO and SRO is reflected in the non-zero value of $\alpha$. For pure MnGe $\alpha$ smoothly increases with temperature and is equal to 0.1 at $T \approx 35$ K. As long as the temperature phase transition is spread over 100 K above $T_N$, these fluctuations could not be related to the typical critical spin fluctuations close to the phase transition to the paramagnetic state. The SRO, or, the helix fluctuations, in MnGe has the clearly thermal origin only at temperatures $T_{N} < T < T_{h}$, as the inverse correlation length $\kappa$ is temperature dependent (Fig.\ref{fig:kappaxt}). The helical fluctuations observed at low temperature range for MnGe could be referred to the same type of the SRO found for the doped compounds at low temperatures. However, the instrumental resolution of this study does not allow to clearly establish this fact. Only at the temperatures below the line corresponding to $\alpha = 0.1$ the structure can be considered as relatively stable one. The temperature evolution of the magnetic structure of Mn$_{1-x}$Fe$_x$Ge with $x = 0.2$ can be discussed similarly to pure MnGe compound.

The situation changes for the compounds with $x > 0.2$. Their temperature evolution was described in the same terms as for the pure MnGe, but the fraction of the stable phase is reduced with $x$ in the range $x \in [0.2 \div 0.4]$ (Fig.\ref{fig:avx}c). The fluctuations observed are considered as the SRO of the quantum nature at temperatures below $T_{QF}$ and of the thermal nature between $T_{QF}$ and $T_{h}$ (Fig.\ref{fig:Temps}). Interesting to note that the temperature $T_{QF}$ and $T_N$ coincide for the compounds with $x = 0.25$ and 0.30. It indicates the correlation between two coexisting fractions. As was shown in \cite{Alt}, the thermal phase transition transition in the MnGe compound is different from already known from Si-based B20 compounds. Its scenario is far from being of the second order. It is realized via the amplification of the fraction of the spiral fluctuations already well below the critical temperature. These fluctuations are gradually replace the stable helical phase upon temperature increase. Here we give the evidence that the change of the Fe-concentration $x$, being the non-thermal parameter, results in the very similar scenario of the phase transition. Thus we use the term "quantum" to emphasize the non-thermal nature of changes in the magnetic structure of the Mn$_{1-x}$Fe$_{x}$Ge compounds.

In accord to \cite{Koretsune_arxiv_2015,Gayles_arxiv_2015,Kikuchi_arxiv_2016} the Fe-replacement of Mn atoms in Mn$_{1-x}$Fe$_{x}$Ge leads to the amplification of the DM interaction. The experiment shows that the competition between RKKY and DMI leads to the destruction of the fragile balance between interactions that built the magnetic order in these compounds. As a result the quantum phase transition from ordered helical structure to the helical SRO is observed at $x_{c} \approx 0.35$. Further increase of $x$ leads to the change of the period of the spin helix $k$ for almost one order of magnitude at $x_{c2} \approx 0.45$. This fact demonstrates the change of the main interaction that built the magnetic helix from the effective RKKY to the DM.

In summary, the comprehensive small-angle neutron scattering study of the temperature evolution of Mn$_{1-x}$Fe$_x$Ge allows one to suggest for consideration the RKKY as the fundamental interaction for helical structure in MnGe. It could be concluded that the order-disorder phase transition at $x_c$ is caused by the modification of the effective Ruderman-Kittel-Kasuya-Yosida exchange interaction within the Heisenberg model of magnetism with $x$ increase. The DMI can be considered as an instrument for destabilization of the ordered helical structure with $x$ or $T$, despite the fact that all Mn$_{1-x}$Fe$_x$Ge compounds crystallizes in B20 type structure. On the other hand, the DMI may be able to break the chiral symmetry of the spiral structure, thus showing another aspect of the coexistence of the different fundamental interactions in these compounds.

The results of this study can be discussed within the context of the Hall effect measurements done for Mn$_{1-x}$Fe$_x$Si \cite{Demishev_PRL_2015} and MnGe \cite{Kanazawa11}. The quantum phase transition in Mn$_{1-x}$Fe$_x$Si is explained as a result of the sign inversion of the ordinary Hall effect, $\rho^N_{yx} = R_0B$, with $x$ \cite{Demishev_PRL_2015}. In case of MnGe, the sign inversion of the topological Hall effect, $\rho^N_{yx} = R_0{\textbf B}^z_{eff}$, occurs as the function of temperature at $T \approx 130$ K \cite{Kanazawa11}. Together with the results of small-angle neutron scattering experiment one can predict that the line marked as $\alpha = 1.0$ in Fig.\ref{fig:Temps} separate the ($T-x$) regions with different signs of the product $\rho^N_{yx} = R_0{\textbf B}^z_{eff}$ at relatively small fields, $H \ll H_{C1}$. Either $R_0$ changes its sign with $x$ at $x_c = 0.35$ or ${\textbf B}^z_{eff}$ changes it sign with $T$ at $T = T_{QF}$. Further Hall effect experiments can prove the validity of our hypothesis. 

The authors are grateful to A. S. Sukhanov for useful discussions. The work was supported by the Russian Foundation of Basic Research (Grant No 14-22-01073, 14-02-00001) and the special program of the Department of Physical Science, Russian Academy of Sciences.

\end{document}